 \documentclass[smallabstract,smallcaptions]{dccpaper}

\usepackage{epsfig}
\usepackage{amsmath}
\usepackage{amssymb}
\usepackage{color}
\usepackage{url}
\usepackage{booktabs}
\usepackage{multirow}
\usepackage{graphicx} 
\usepackage{subcaption} 

\newlength{\figurewidth}
\newlength{\smallfigurewidth}

\begin{document}

\title
{\large
\textbf{Generative Preprocessing for Image Compression with Pre-trained Diffusion Models}
}

\author{%
Mengxi Guo, Shijie Zhao$^{\ast}$\thanks{$^{\ast}$Corresponding author}, Junlin Li, Li Zhang\\[0.5em] 
{\small\begin{minipage}{\linewidth}\begin{center}
\begin{tabular}{c}
Bytedance Inc., Shenzhen, China \\
Bytedance Inc., San Diego, CA, 92122 USA \\
\{guomengxi.qolab, zhaoshijie.0526, lijunlin.li, lizhang.idm\}@bytedance.com
\end{tabular}
\end{center}\end{minipage}}
}


\maketitle
\thispagestyle{empty}

\begin{abstract}
Preprocessing is a well-established technique for optimizing compression, yet existing methods are predominantly Rate-Distortion (R-D) optimized and constrained by pixel-level fidelity. This work pioneers a shift towards Rate-Perception (R-P) optimization by, for the first time, adapting a large-scale pre-trained diffusion model for compression preprocessing. We propose a two-stage framework: first, we distill the multi-step Stable Diffusion 2.1 into a compact, one-step image-to-image model using Consistent Score Identity Distillation (CiD). Second, we perform a parameter-efficient fine-tuning of the distilled model's attention modules, guided by a Rate-Perception loss and a differentiable codec surrogate. Our method seamlessly integrates with standard codecs without any modification and leverages the model's powerful generative priors to enhance texture and mitigate artifacts. Experiments show substantial R-P gains, achieving up to a 30.13\% BD-rate reduction in DISTS on the Kodak dataset and delivering superior subjective visual quality.
\end{abstract}

\section*{Introduction}

Preprocessing is a ubiquitous technique for optimizing compression, capable of yielding significant Rate-Distortion (R-D) gains for both image and video codecs. In recent years, the proliferation of deep learning in the compression domain has led to a surge in learning-based preprocessing methods. These efforts can be broadly categorized into two main streams: those that enhance the R-D performance of the compression algorithm itself, and those that improve the performance of downstream machine vision tasks on the compressed data.

Within the first category, a common paradigm involves employing a multi-layer convolutional neural network (CNN) as the preprocessing module. This module is typically trained end-to-end with a differentiable surrogate of a traditional codec, guided by R-D optimization. Despite their success, these methods suffer from two primary limitations. First, they heavily rely on pixel-level loss functions (e.g., MSE or L1) as the primary optimization objective~\cite{prepjpeg, bap}, while losses that correlate better with subjective quality, such as adversarial or perceptual losses~\cite{gan}, are often relegated to a supplementary role. While necessary to prevent drastic image alterations, this strategy forces a difficult trade-off between rate, distortion, and perception. This constraint, explained by the R-D-P trade-off~\cite{blau2019rethinking}, inherently limits the full potential of preprocessing. Second, existing works typically employ small models trained from scratch, failing to leverage the power of modern, large-scale pre-trained foundation models~\cite{sd2.1, dalle, imagen} that have shown superior performance on various generative and restorative tasks. However, these models' application to the image compression preprocessing task remains unexplored.

Motivated by these observations, we are the first work to introduce a pre-trained foundation model for the task of image compression preprocessing. The state-of-the-art pretrained models are predominantly Text-to-Image (T2I) diffusion models~\cite{dalle, imagen, sd2.1, DiT}. However, a fundamental misalignment exists between the objectives of T2I synthesis and compression preprocessing, primarily concerning the prohibitively large model size and slow inference speed of T2I models, as well as their often-limited detail fidelity. To address these issues, we propose a novel generative preprocessing framework. First, to create a lightweight and efficient model, we distill a large pre-trained T2I model using Variational Score Distillation (VSD)~\cite{gendr} to obtain a compact one-step model. This distilled model has significantly fewer parameters and lower computational overhead, although with intentionally reduced generative capabilities.

Subsequently, we fine-tuned the distilled model to adapt it to the compression preprocessing task. Given that most state-of-the-art generative models are built upon the transformer block in UNet, we focus our investigation on how to efficiently fine-tune such structures. We experimentally find that by updating only a small fraction of the model's parameters, we can instill the ability for compression optimization without sacrificing the rich semantic understanding and texture synthesis capabilities inherited from the original pre-trained model. During fine-tuning, we design a differentiable BGP (diff-BPG) to simulate the distortion and rate fluctuations of a real-world codec, and employ a Rate-Perception loss function to manage the trade-off between processing intensity and bitrate. Finally, compared to existing CNN-based preprocessing methods, our generative approach not only achieves a BD-rate reduction of 30.13\% on the Kodak dataset in objective metrics but also demonstrates superior performance in subjective evaluations.

\section*{Methods}

\begin{figure*}[t!]
    \centering
    \includegraphics[width=\textwidth]{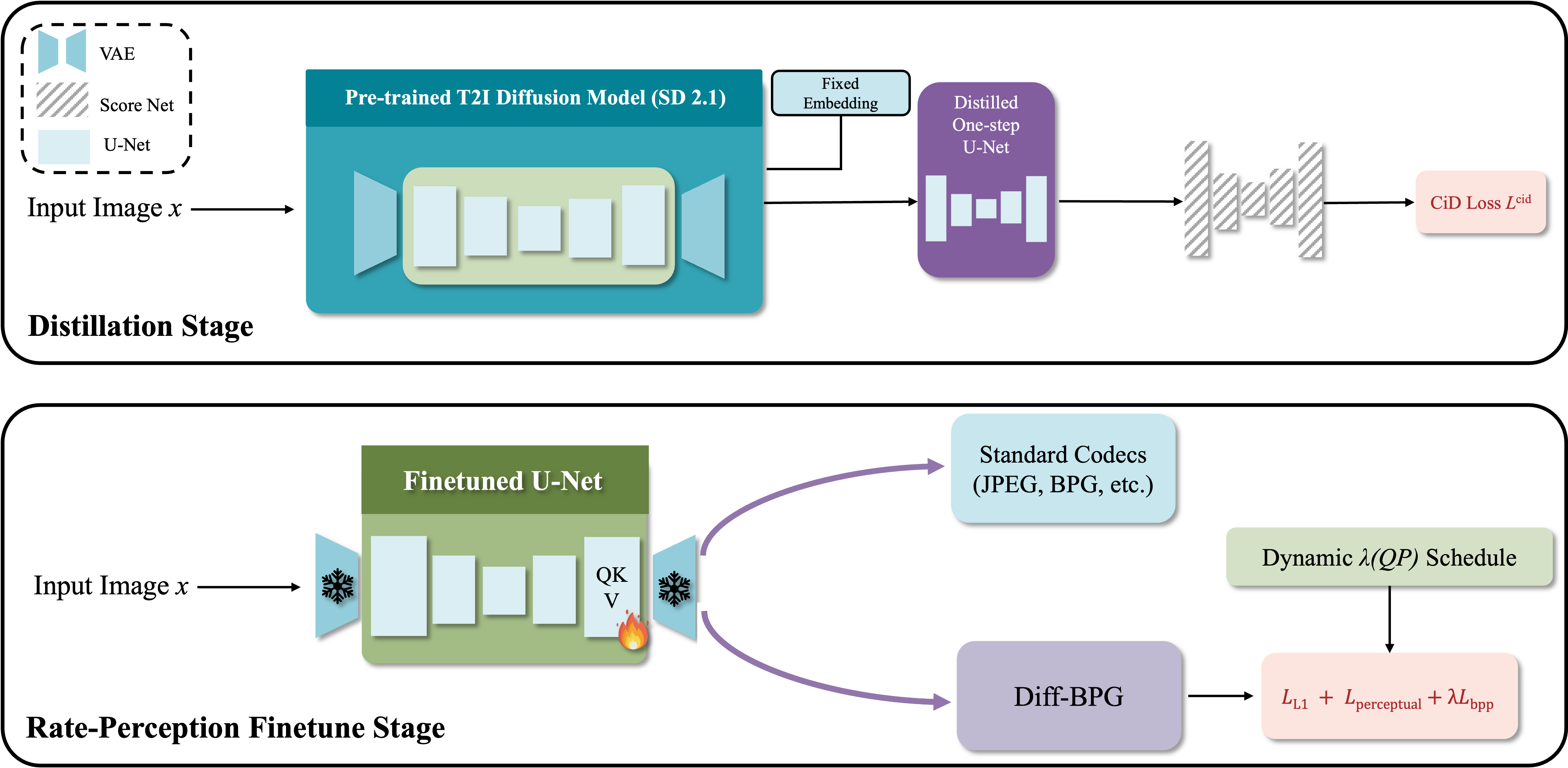}
    \caption{An overview of our proposed two-stage framework. \textbf{(Top) Distillation Stage:} We distill a pre-trained, multi-step Stable Diffusion 2.1 model into a compact, one-step U-Net using Consistent Score Identity Distillation (CiD). The VAE and text conditioning (replaced by a fixed embedding) are kept frozen. \textbf{(Bottom) Rate-Perception Finetune Stage:} The distilled one-step generator is fine-tuned for the preprocessing task. An input image is processed by the finetuned U-Net (with frozen VAE) and then passed through a differentiable BPG (Diff-BPG) surrogate. A composite loss, balancing L1, perceptual, and bitrate terms with a dynamic QP-based schedule, guides the optimization of the U-Net's attention modules.}
    \label{fig:method_overview}
\end{figure*}

The overall architecture of our proposed method, illustrated in Figure~\ref{fig:method_overview}, consists of two main stages: a Distillation Stage to create a compact one-step generator, and a Rate-Perception Finetune Stage to adapt the generator for the compression preprocessing task.

\subsection*{Diffusion Models}

Our methodology is built on a pre-trained Stable Diffusion  2.1~\cite{sd2.1}, which has demonstrated state-of-the-art performance in text-to-image synthesis. Diffusion models are a class of probabilistic generative models that learn to generate data by reversing a gradual noising process. This process consists of two main stages: a forward diffusion process and a reverse denoising process.

The forward process, $q$, gradually adds Gaussian noise to an initial data sample $x_0$ over a sequence of $T$ timesteps. At each timestep $t$, the data $x_t$ is produced by sampling from a Gaussian distribution conditioned on the previous sample $x_{t-1}$:
\begin{equation}
q(x_t | x_{t-1}) = \mathcal{N}(x_t; \sqrt{1 - \beta_t} x_{t-1}, \beta_t \mathbf{I})
\end{equation}
where $\{\beta_t\}_{t=1}^T$ is a predefined variance schedule.

The reverse process aims to learn the data distribution by reversing the forward noising process. This is achieved by training a neural network, typically a U-Net architecture denoted as $\epsilon_\theta$, to predict the noise $\epsilon$ that was added to the noisy data $x_t$ at timestep $t$.

To improve computational efficiency, our work operates in a lower-dimensional latent space. An autoencoder, consisting of an encoder $\mathcal{E}$ and a decoder $\mathcal{D}$, is pre-trained to compress an image $x$ into a latent representation $z = \mathcal{E}(x)$. The diffusion process described above is then applied to this latent variable $z$.

\begin{figure*}[t!]
    \centering
    \includegraphics[width=0.8\textwidth]{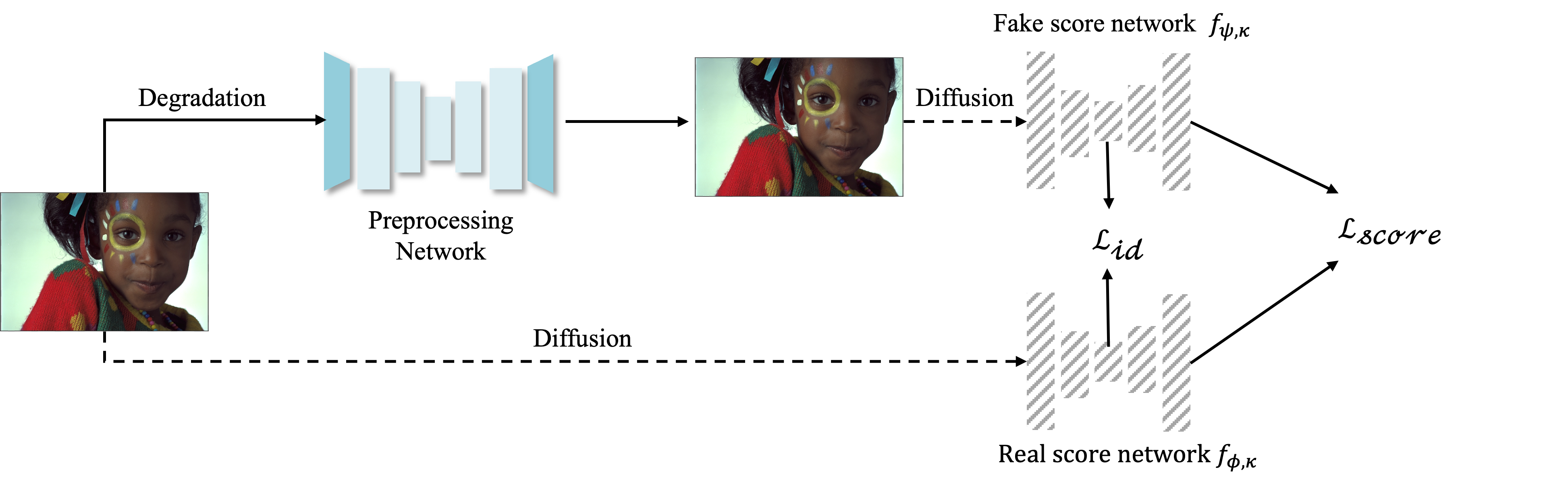}
    \caption{The Detail of the Distillation Stage. }
    \label{fig:distillation_diagram}
\end{figure*}

\subsection*{Distillation}

The iterative nature of the standard Stable Diffusion 2.1 model imposes a prohibitive computational cost for compression preprocessing. To create an efficient one-step generator, we must adapt the distillation process itself. A standard noise-to-image distillation is ill-suited for our preprocessing task, as it overlooks the strong input-output correlation inherent in image-to-image translation. We therefore adapt Consistent Score Identity Distillation (CiD)~\cite{gendr}, a framework designed for image-to-image tasks, to reframe the distillation as a guided image translation problem.

The CiD framework, visually detailed in Figure~\ref{fig:distillation_diagram}, is built to address the instability in Variational Score Distillation (VSD), which relies on a generated latent $z_g$ that has fluctuating quality. CiD stabilizes training by using a pre-trained teacher model, which acts as a \textbf{real score network}, to guide a student model, or \textbf{fake score network}. Crucially, it also introduces a high-fidelity latent anchor, which we define as the latent representation $z_h$ of the original, unprocessed image. The final training objective is a combination of two core loss terms derived from this concept.

The first term is the \textbf{identity loss}, $\mathcal{L}_{\text{id}}$, which is the primary driver of CiD. It replaces the unstable target $z_g$ with the stable anchor $z_h$. This loss ensures consistency by aligning the score difference between the real score network $f_{\phi,\kappa}$ and the fake score network $f_{\psi,\kappa}$ with the direction towards the high-fidelity anchor $z_h$:
\begin{equation}
    \mathcal{L}_{\text{id}} = \mathbb{E}_{t, \epsilon, c, z_t} \left[ w(t) \left\langle f_{\phi, \kappa}(z_t, t, c) - f_{\psi, \kappa}(z_t, t, c), f_{\phi, \kappa}(z_t, t, c) - z_h \right\rangle \right].
\end{equation}
The second term is a \textbf{score difference loss}, $\mathcal{L}_{\text{score}}$, which functions similarly to the original VSD objective but is formulated to circumvent the inability to compute the fake score network's gradient $\nabla_\theta \psi$. Let's denote it as $\mathcal{L}_{1}$ following prior work. These two losses are then combined with an empirical weighting factor $\xi$ to form the final CiD objective:
\begin{equation}
    \mathcal{L}_{\text{cid}} = \mathcal{L}_{\text{id}} - \xi \mathcal{L}_{\text{score}}.
\end{equation}
This distillation framework, which leverages both a powerful generative prior and a strong image-conditional constraint, yields a highly efficient one-step model suitable for our task. The final architecture comprises a compact U-Net and the original VAE, with the text encoder discarded and replaced by a static embedding, reducing inference time to approximately 1\% of the original model.

\subsection*{Rate-Perception Finetune}

Although the distilled model is computationally efficient, it still operates as a generative model and cannot be directly applied to compression preprocessing. Therefore, we introduce a fine-tuning stage to adapt its capabilities from generation to preprocessing, shown in the bottom panel of Figure~\ref{fig:method_overview}. The core architecture of the diffusion model consists of a U-Net and a VAE. In line with standard practices where the VAE is used as a fixed feature extractor, we freeze the pre-trained VAE weights and focus exclusively on fine-tuning the U-Net.

Given the U-Net's attention-based architecture, we adopt a parameter-efficient fine-tuning strategy. We discovered that updating only the parameters of the attention modules within each Transformer block is sufficient to achieve excellent performance. Specifically, for a given attention module, the parameters consist of the projection matrices for query ($W_Q$), key ($W_K$), and value ($W_V$). The attention mechanism is formulated as:
\begin{equation}
\text{Attention}(Q, K, V) = \text{softmax}\left(\frac{QK^T}{\sqrt{d_k}}\right)V,
\end{equation}
where $Q=XW_Q$, $K=XW_K$, $V=XW_V$ are the projected query, key, and value matrices from the input sequence $X$. By only updating these projection matrices, we allow the model to learn the task-specific mapping required for preprocessing while preserving the rich, pre-trained features from the foundational model.

\paragraph{Differentiable BPG Codec}
To enable end-to-end training, we require a differentiable surrogate for the target codec. Inspired by ~\cite{zhao2024preprocessing}, we design a differentiable BPG (diff-BPG) module that simulates the key stages of BPG compression. To reduce spatial redundancy, our simulation incorporates an intra-prediction step which utilizes a set of candidate modes (e.g., planar, DC, angular) based on neighboring pixels. The selection of the best prediction mode, a non-differentiable $\text{argmin}$ operation over the candidates' Rate-Distortion costs, is relaxed using a soft-argmin approximation to enable gradient propagation~\cite{soft-argmin}. The other major challenge, the non-differentiable rounding operation within quantization, is overcome by employing a Fourier series-based approximation:
\begin{equation}
\hat{y} = y - \frac{1}{2\pi} \sum_{n=1}^{N} \frac{(-1)^{n+1}}{n} \sin(2\pi n y),
\end{equation}
where $y$ is the value after quantization division and $\hat{y}$ is its differentiable, rounded counterpart. Finally, the entropy coding stage is replaced with a learned entropy model~\cite{balle2018variational} that provides a differentiable estimate of the bits per pixel (bpp).

\paragraph{Loss Function}
Our fine-tuning process is guided by a composite loss function designed to balance rate and perception. The total loss $\mathcal{L}_{\text{total}}$ is defined as:
\begin{equation}
\mathcal{L}_{\text{total}} = \mathcal{L}_{\text{L1}} + w_p \mathcal{L}_{\text{perceptual}} + \lambda \mathcal{L}_{\text{bpp}},
\end{equation}
where $\mathcal{L}_{\text{L1}}$ is the L1 loss between the original and the preprocessed-reconstructed image, which helps preserve fine-grained details. $\mathcal{L}_{\text{perceptual}}$ is the DISTS loss~\cite{dists}, a metric known to correlate well with human perceptual quality. $\mathcal{L}_{\text{bpp}}$ is the estimated bitrate from our diff-BPG module. The term $w_p$ is a fixed weight for the perceptual loss. The hyperparameter $\lambda$ controls the trade-off between perception and rate. We dynamically adjust $\lambda$ based on the Quantization Parameter (QP) used during training, following an empirically derived schedule:
\begin{equation}
\lambda(\text{QP}) = e^{(w_1 \cdot \text{QP} + w_2)},
\end{equation}
where we set $w_1 = -0.115$ and $w_2 = 1.145$. This allows the model to learn an effective R-P trade-off across a wide range of compression levels.

\begin{table*}[t!]
\centering
\caption{BDBR (\%) reduction compared to the anchor codec without preprocessing. Lower values are better, indicating greater bitrate savings for the same perceptual quality. Our method consistently outperforms the TDP baseline across all codecs and datasets.}
\label{tab:bdbr_results}
\resizebox{0.8\textwidth}{!}{%
\begin{tabular}{@{}llllcccccc@{}} 
\toprule
\multirow{2}{*}{\textbf{Dataset}} & \multirow{2}{*}{\textbf{Metric}} & \multicolumn{2}{c}{\textbf{JPEG}} & \multicolumn{2}{c}{\textbf{WebP}} & \multicolumn{2}{c}{\textbf{BPG}} \\ 
\cmidrule(lr){3-4} \cmidrule(lr){5-6} \cmidrule(lr){7-8} 
 &  & \textbf{TDP} & \textbf{Ours} & \textbf{TDP} & \textbf{Ours} & \textbf{TDP} & \textbf{Ours} \\ \midrule
\multirow{3}{*}{\textbf{CLIC}} & LPIPS & -1.62 & \textbf{-5.74} & -3.65 & \textbf{-8.30} & -11.18 & \textbf{-23.16} \\
 & DISTS & -3.99 & \textbf{-9.74} & -7.88 & \textbf{-21.43} & -11.2 & \textbf{-27.68} \\
 & TOPIQ-fr & -0.23 & \textbf{-7.88} & -6.25 & \textbf{-23.80} & -6.48 & \textbf{-22.38} \\ \midrule
\multirow{3}{*}{\textbf{Kodak}} & LPIPS & -1.33 & \textbf{-6.14} & -3.37 & \textbf{-7.43} & -6.17 & \textbf{-12.57} \\
 & DISTS & -4.27 & \textbf{-13.69} & -8.30 & \textbf{-25.22} & -12.62 & \textbf{-30.13} \\
 & TOPIQ-fr & -2.94 & \textbf{-7.22} &  -1.61 & \textbf{-16.40} & -7.27 & \textbf{-15.64} \\ \bottomrule
\end{tabular}%
}
\end{table*}

\section*{Experiments}

\subsection*{Implementation Details}

Our training process is divided into two distinct stages: distillation and fine-tuning. For the distillation stage, we initialize our model from the official SD2.1. The distillation is performed on a combined dataset comprising LSDIR~\cite{lsdir} and FFHQ~\cite{ffhq}. We use a fixed learning rate of $1 \times 10^{-5}$ throughout this stage. In the subsequent Rate-Perception fine-tuning stage, we use the DIV2K and Flickr2K~\cite{flickr2k} datasets as our primary training data, from which we extract random $256 \times 256$ patches. 

To enhance the model's focus on challenging image regions, we introduce a data curation step based on Just Noticeable Distortion (JND)~\cite{jnd}. We filter the patches and retain only those with a JND score greater than 0.8. This strategy prioritizes patches with high texture complexity, such as backgrounds, where the human visual system is less sensitive to minor alterations and where our generative preprocessing can provide the most benefit. During fine-tuning, our diff-BPG module is subjected to a randomly fluctuating Quantization Parameter (QP) within the range of $[10, 50]$. The fine-tuning is optimized using the Adam optimizer with an initial learning rate of $1 \times 10^{-3}$. We employ a cosine annealing schedule to decay the learning rate down to a final value of $1 \times 10^{-8}$. Our entire framework is implemented in PyTorch with CUDA support. All experiments were conducted on a server equipped with 8 NVIDIA V100 GPUs.

For evaluation, we utilize standard benchmark datasets: the Kodak dataset and the CLIC Professional Validation dataset. In the testing phase, we evaluate the performance of our preprocessed images using several real-world image compression codecs: JPEG, WebP, and BPG. We utilize bits per pixel (bpp) as the measure to gauge the coding cost incurred during the compression process. To assess perceptual quality, we employ a suite of well-established metrics, including LPIPS~\cite{lpips}, DISTS~\cite{dists}, and TOPIQ-fr~\cite{topiq}. The final performance gain is comprehensively quantified by calculating the Bjontegaard Delta Bit Rate (BDBR), which measures the average bitrate savings for the same perceptual quality.

\subsection*{Experimental Results}

In this section, we present a comprehensive evaluation of our proposed generative preprocessing framework. We conduct extensive quantitative and qualitative comparisons against two baselines: the standard codec without any preprocessing (referred to as the anchor) and a state-of-the-art preprocessing method, TDP~\cite{bap}.

\paragraph{BDBR Performance.}
To quantify the performance gains, we calculate the Bjontegaard Delta Bit Rate (BDBR) savings across three widely-used perceptual metrics: LPIPS, DISTS, and TOPIQ-fr. The results, presented in Table~\ref{tab:bdbr_results}, demonstrate the superiority of our method. Across both the Kodak and CLIC validation datasets, our approach consistently achieves substantially higher bitrate reductions compared to the TDP baseline for all evaluated codecs. For example, when paired with the BPG codec, our method achieves remarkable DISTS BDBR savings of \textbf{-27.68\%} on CLIC and \textbf{-30.13\%} on Kodak, outperforming TDP. These large and consistent improvements underscore the significant potential of leveraging pre-trained generative models for perception-oriented compression preprocessing.

\begin{figure*}[t!]
    \centering
    \begin{subfigure}[b]{0.32\textwidth}
        \includegraphics[width=\textwidth]{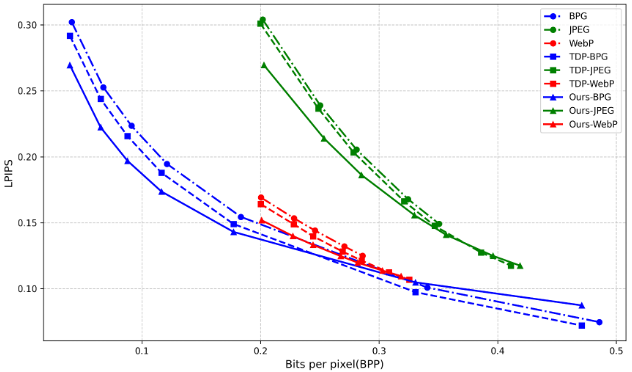}
        \caption{CLIC - LPIPS}
        \label{fig:clic_lpips_rp}
    \end{subfigure}
    \hfill 
    \begin{subfigure}[b]{0.32\textwidth}
        \includegraphics[width=\textwidth]{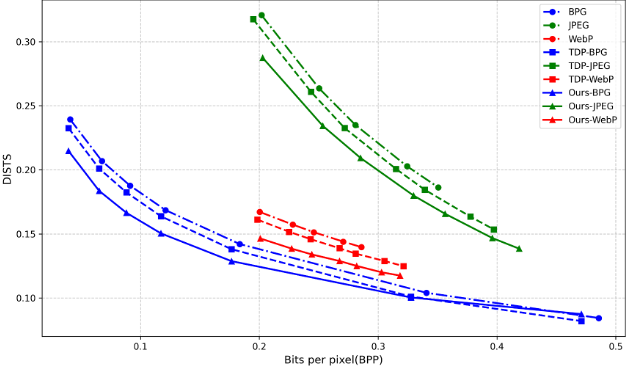}
        \caption{CLIC - DISTS}
        \label{fig:clic_dists_rp}
    \end{subfigure}
    \hfill 
    \begin{subfigure}[b]{0.32\textwidth}
        \includegraphics[width=\textwidth]{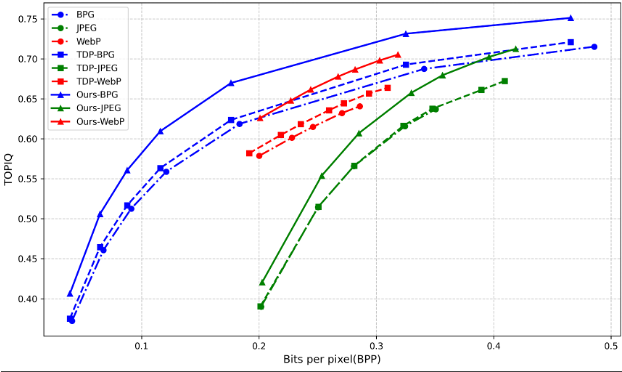}
        \caption{CLIC - TOPIQ-fr}
        \label{fig:clic_topiq_rp}
    \end{subfigure}

    \vspace{0.4cm} 

    \begin{subfigure}[b]{0.32\textwidth}
        \includegraphics[width=\textwidth]{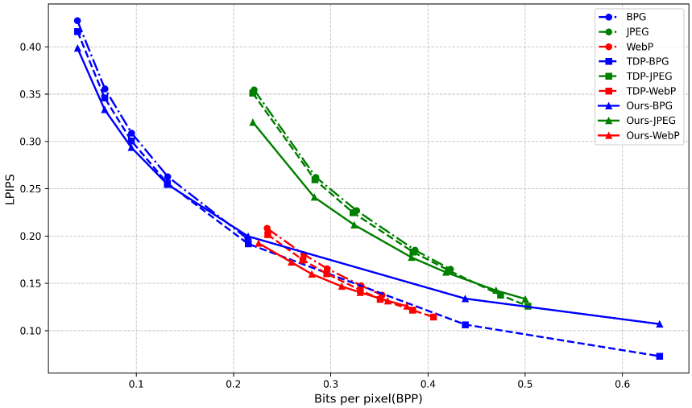}
        \caption{Kodak - LPIPS}
        \label{fig:kodak_lpips_rp}
    \end{subfigure}
    \hfill 
    \begin{subfigure}[b]{0.32\textwidth}
        \includegraphics[width=\textwidth]{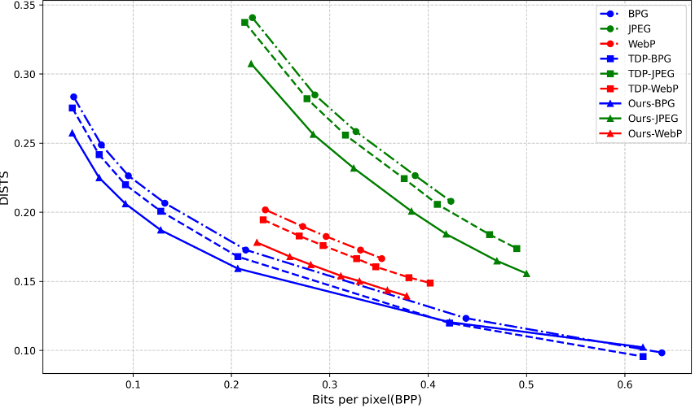}
        \caption{Kodak - DISTS}
        \label{fig:kodak_dists_rp}
    \end{subfigure}
    \hfill 
    \begin{subfigure}[b]{0.32\textwidth}
        \includegraphics[width=\textwidth]{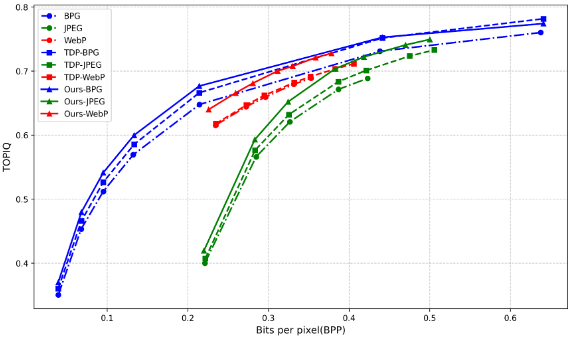}
        \caption{Kodak - TOPIQ-fr}
        \label{fig:kodak_topiq_rp}
    \end{subfigure}
    
    \caption{Rate-Perception (R-P) curves on the CLIC validation dataset and Kodak dataset. These plots compare the perceived quality (LPIPS, DISTS, TOPIQ-fr) against the bitrate (bpp) for different codecs (JPEG, WebP, BPG) and preprocessing methods (TDP, Ours). Our method consistently achieves better perceptual quality at lower bitrates across all metrics and datasets.}
    \label{fig:rp_curves_combined}
\end{figure*}

\paragraph{Rate-Perception Curves.}
For a more detailed analysis, we plot the Rate-Perception (R-P) curves for all methods on both datasets, as shown in Figure~\ref{fig:rp_curves_combined}. The curves visualize the trade-off between perceptual quality and bitrate , where a curve positioned higher and to the left indicates superior performance. In the crucial low-to-mid bitrate range (e.g., $bpp < 0.5$), our method's curves consistently dominate those of both the anchor and the TDP baseline across all metrics and datasets. This demonstrates that our generative approach delivers significantly better perceptual quality at the same or even lower bitrates, which aligns with the strong BDBR results.

However, we observe a performance crossover at higher bitrates (e.g., $bpp > 0.5$), where our method is sometimes outperformed by TDP or the anchor. This phenomenon is an inherent trade-off of our model's generative nature. Because our model is fine-tuned from a generative foundation, it is optimized to synthesize plausible and perceptually pleasing textures, which can involve modifying the original image content more significantly than traditional methods. At high bitrates where the anchor codec can already preserve most details, the modifications made by our model, while perceptually realistic, may deviate more from the original source, leading to a penalty from similarity-based perceptual metrics like LPIPS and DISTS. This trade-off, where generative methods excel in low-bitrate regimes by creating realistic details, is a known characteristic and has been observed in previous generative compression work~\cite{hific}.

\begin{figure*}[t!]
    \centering
    \includegraphics[width=\textwidth]{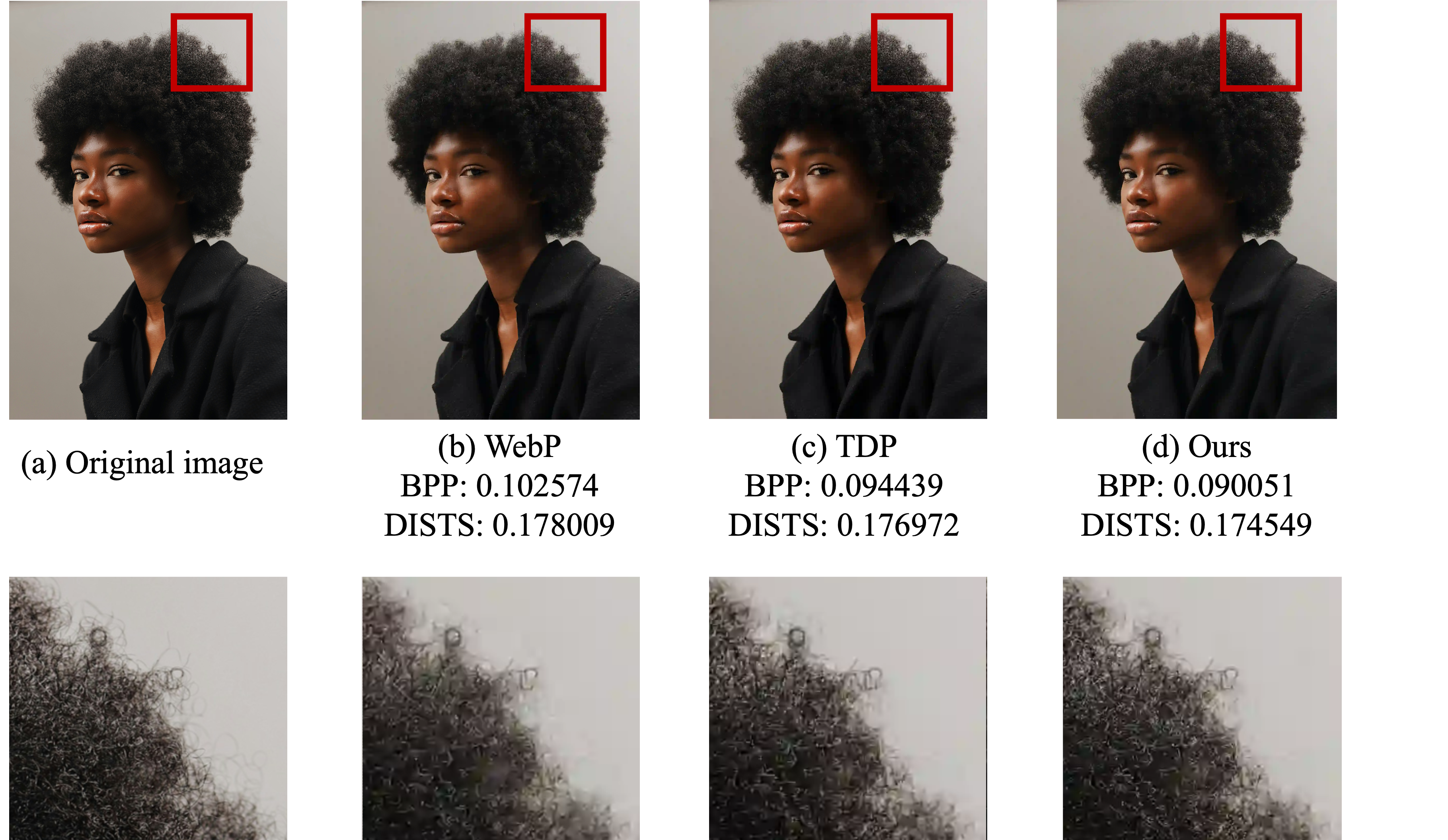}
    \caption{Qualitative comparison on a sample from the CLIC validation dataset. Our method effectively removes compression artifacts while preserving and enhancing fine-grained textures, leading to superior visual quality compared to the anchor WebP and TDP.}
    \label{fig:qualitative_comparison}
\end{figure*}

\paragraph{Qualitative Comparison}
Beyond quantitative metrics, we provide qualitative comparisons to visually illustrate the advantages of our generative preprocessing approach. As shown in Figure~\ref{fig:qualitative_comparison}, we present zoomed-in regions of a representative image from the CLIC validation dataset compressed to a similar bitrate. The image compressed directly with WebP exhibits noticeable artifacts, such as blocking and blurring, especially in textured areas. While TDP alleviates some of these artifacts, it tends to over-smooth fine details, resulting in a loss of natural texture. In contrast, our method effectively removes compression artifacts while simultaneously restoring and enhancing plausible high-frequency details. The rich texture synthesis capability inherited from the pre-trained generative model allows our method to produce a visually pleasing result that is both sharp and natural, demonstrating a clear perceptual advantage over the competing methods.

\section*{Conclusion}

In this paper, we introduced a novel framework for rate-perception-optimized compression preprocessing, marking the first successful adaptation of a large-scale pre-trained generative model for this task. Our approach, combining efficient model distillation with a targeted Rate-Perception fine-tuning strategy, effectively bridges the gap between generative models and traditional compression pipelines. Experiments demonstrate that our method delivers superior visual quality and achieves significant bitrate savings with a better perceptual score.



\Section{References}
\bibliographystyle{IEEEtran}
\bibliography{refs}

\end{document}